# Photosensitivity at 1550 nm and Bragg grating inscription in As$_2$Se$_3$ microwires

Raja Ahmad* and Martin Rochette

*Department of Electrical and Computer Engineering, McGill University, Montreal (QC), Canada, H3A 2A7*

*Corresponding author: raja.ahmad@mail.mcgill.ca

**Abstract**

We report the first experimental observation of photosensitivity in As$_2$Se$_3$ glass at a wavelength of 1550 nm. We utilize this photosensitivity to induce the first Bragg gratings using a laser source at a wavelength of 1550 nm. We quantify the photosensitivity thresholds related to exposition intensity and exposition time. Finally, we demonstrate that As$_2$Se$_3$ Bragg gratings are widely tunable in wavelength as the microwire can withstand an applied longitudinal strain of $4 \times 10^4$ µε.

Keywords: Chalcogenide; Arsenic triselenide As$_2$Se$_3$; Photosensitivity; Microwires; Bragg gratings

PACS: 42.70.Gi; 42.79.Dj; 42.70.Ce

In recent years, chalcogenide glasses have attracted a lot of interest for applications such as sensing, mid-infrared light transmission and high data rate signal processing [1]. This interest of chalcogenide glasses is attributed to their large photosensitivity, large nonlinear coefficient and a wide transmission window covering the telecommunication band around 1.55 µm and up to 20 µm in the mid-infrared [2]. In the past, the photosensitivity of chalcogenide glasses has been utilized to write Bragg gratings in fibers as well as in waveguides [3, 4]. Recently, we have reported the first side-written Bragg gratings in As$_2$Se$_3$ microwires [5]. However, in all of these cases, light at a wavelength (λ) corresponding to a photon energy ($E_v$) equal or close to the material bandgap ($E_v$ = 1.9 eV or λ=650 nm for $As_2Se_3$ and $E_v$=2.4 eV or λ=820 nm for $As_2S_3$) was used to realize a refractive index modulation that induces a Bragg grating. Up to now, the photosensitivity in $As_2S_3$ glass has been studied and reported for low photon energy with respect to the bandgap energy i.e., at a wavelength of 1550 nm [6, 7], but yet there has been no report on the photosensitivity of $As_2Se_3$ in this wavelength range.

In this letter, we report the first experimental observation of $As_2Se_3$ photosensitivity at a wavelength of 1550 nm. We use the photosensitivity to write Bragg gratings in $As_2Se_3$ microwires following the Hill's approach [8]. The grating formation is monitored during the photo-exposure to reveal the time evolution of the induced refractive index change. This allows quantifying the intensity and time thresholds for inducing a significant change in the refractive index of $As_2Se_3$. The tunability of the Bragg grating with applied longitudinal strain is also measured.

Microwires with a waist diameter of 1 μm and a length of 4 cm were prepared from index-guiding $As_2Se_3$ fibers (Coractive High-Tech inc.) using the process detailed in [9]. Fig. 1 shows the all-fiber interferometric setup used to photo-induce the Bragg gratings with laser sources at a wavelength of 1550 nm. Two distinct approaches were used to write the Bragg gratings. In the first approach (A), the output of a continuous-wave (CW) laser was carved into pulses of 1 ns in duration and a repetition rate of 131.072 kHz. The modulation was provided from two cascaded intensity modulators driven by a pulse pattern generator. The modulated signal had an extinction ratio of 40 dB. This large extinction ratio is important for keeping a low average power while maintaining a relatively high peak power, so as to minimize the thermal effects due to absorption in the chalcogenide glass. After modulation, the peak power of the pulses was increased above the photosensitivity threshold using an erbium doped fiber amplifier (EDFA). An optical bandpass filter (BPF 1) was used to eliminate the amplified spontaneous emission (ASE) noise added from the EDFA. In a second approach (B), the output of a mode-locked laser with femtosecond pulses and a repetition rate of 20 MHz was passed through a ~0.25 nm bandpass filter (BPF 2), which led to pulses with a full-width at half-maximum of ~22 ps. In both scenarios (A) and (B), the pulsed signal was sent through an optical attenuator followed by a 99:1 fiber coupler, with the 1% signal sent to a power meter while the 99% signal was sent to a 50:50 fiber coupler. The attenuator and 99:1 coupler assembly allowed monitoring and controlling the amount of power sent into the grating fabrication part of the setup. The two output ports of the 50:50 coupler were then connected to each end of the microwire. The optical signals interfered inside the $As_2Se_3$ microwire, thereby creating a standing wave pattern and photo-inducing a periodic modulation of the refractive index, with a period equal to half the signal wavelength. This resulted in the formation of a Bragg grating inside the uniform waist region of the microwire. The Bragg grating was formed only in the uniform waist region because the mode intensity was considerably enhanced and thus exceeded the threshold value earlier, as compared to that in the transition and/or unstretched region of the microwire [5]. In order to observe the evolution of the grating, a 90:10 coupler was inserted at each end of the microwire, and the transmission of a broadband signal through the microwire was observed on an optical spectrum analyzer (OSA). Figs. 2 (a) and (b) show the spectra for the gratings formed by using the pulsed signal generated by using the approaches A and B, respectively. The inset in Fig. 2(a) shows the 1-ns square signal trace on the oscilloscope and the inset of Fig. 2(b) contains the autocorrelation trace of the 22 ps Gaussian pulses.

Fig. 3 shows the time evolution of a typical Bragg grating. A transmission dip appeared at the input signal wavelength after ~ 5 minutes of time exposure and reached -6 dB after 10 minutes of time exposure. After this, there was a rapid decrease in the transmission dip as it reached less than -30 dB within the next 3 minutes. Beyond this time, the exact level of the transmission dip was masked by the noise floor of the OSA and thus, could not be resolved accurately. We could, however, observe the increase in width of the grating resonance which allowed the calculation of the refractive index change, by fitting the spectrum with coupled mode theory. Importantly, the shift in Bragg wavelength towards the longer wavelengths during the process of grating formation revealed that the refractive index of $As_2Se_3$ glass increases upon exposure to 1550 nm light. This contrasts with our previous findings with the photo-inducing wavelength of 633 nm, where the refractive index was instead observed to decrease upon photo-exposure [5].

We also performed experiments to determine the intensity and time thresholds required to induce a significant refractive index change. In the first set of experiments, we exposed each microwire sample with a different input intensity for a period of 20 minutes and observed the induced refractive index change. From Fig. 4(a), three regions of the intensity range can be identified. At < 10 MW/cm$^2$ (region I), there was no grating formation, even after > 90 minutes of photo-exposure (not in figure). Take note that the intensity provided is the sum of the pulse intensities circulating in both directions in the microwire. For the sum intensity values in the range of 10-225 MW/cm$^2$ (region II), the induced refractive index change followed a linear increasing function. The threshold intensity was 333 MW/cm$^2$ - or effectively 666 MW/cm$^2$, which is the intensity of the standing wave - beyond which the increase in refractive index change was strongly enhanced (region III). The threshold intensity was defined as the point where the linear interpolation curves in the regions II and III intersect. We also characterized the time threshold for the formation of a Bragg grating. This was accomplished by monitoring the grating evolution over time with a fixed input intensity. For a input intensity sum equal to 425 MW/cm$^2$, we measured a time threshold of 8.2 minutes, again estimated from the intersection of the linear interpolation curves in the two regions of time scale. A time threshold is associated to a given intensity value and changes upon varying the intensity. An important observation is that a time threshold was attained only when the intensity value lied in the region III of Fig. 4(a), otherwise the refractive index change increased linearly up to a certain amount time and then remained constant. This was observed by exposing a sample with sum intensity value 52 MW/cm$^2$ [in region II of Fig. 4(a)] for the time duration even longer than 2 hours. The evolution of the refractive index change is shown as inset in the Fig. 4 (b).

Finally, the shift in Bragg wavelength due to the applied longitudinal strain ε was quantified. A Bragg grating was written in a 1 μm diameter wire using the approach A, as discussed above. The total length of the uniform section [5, 8] of the microwire was 4 cm. In order to perform the strain measurements, each end of the microwire was fixed on a motorized translation stage. One end of the microwire was pulled slowly while the corresponding shift in Bragg wavelength was recorded. Fig. 5 shows that the Bragg wavelength shifts as a function of applied strain, fitting with a linear function with a correlation coefficient of $R^2 = 0.9998$. A sensitivity of approximately 1 pm/με is obtained from the linear fit, which agrees well with the theoretical calculations [10, 11]. Wavelength shift measurements over a spectral range of 1550-1564 nm are provided in Fig. 5, with the spectra provided as inset in the figure. Wavelength shift measurement was however limited by the spectral width of the broadband ASE source used for the characterization. The microwire could be stretched over a length of > 1.6 mm - corresponding to a strain value of $4 \times 10^4$ με - which shows that the microwires are remarkably strong and extensible. In comparison, a maximum strain of ~$10^4$ με can be applied to silica fiber Bragg gratings before the fiber breaks or its response becomes nonlinear [12]. By extrapolating the linear fit in Fig. 5, this strain value corresponds to a wavelength shift of > 43 nm. We therefore, conclude that the grating can potentially be tuned over a wide wavelength range of 1550-1593 nm.

In conclusion, we have characterized the photosensitivity in As$_2$Se$_3$ at a wavelength of 1550 nm. We utilized this photosensitivity to make Bragg gratings in As$_2$Se$_3$ microwires and determined the intensity and time thresholds of the photosensitivity process. Finally, Bragg gratings written in the hybrid microwires can experience a strain of $4 \times 10^4$ με, corresponding to a wavelength tunability range from 1550 nm up to 1593 nm.

This work was supported by the Fonds Québecois pour la Recherche sur la Nature et les Technologies (FQRNT).

# FIGURES and FIGURE CAPTIONS

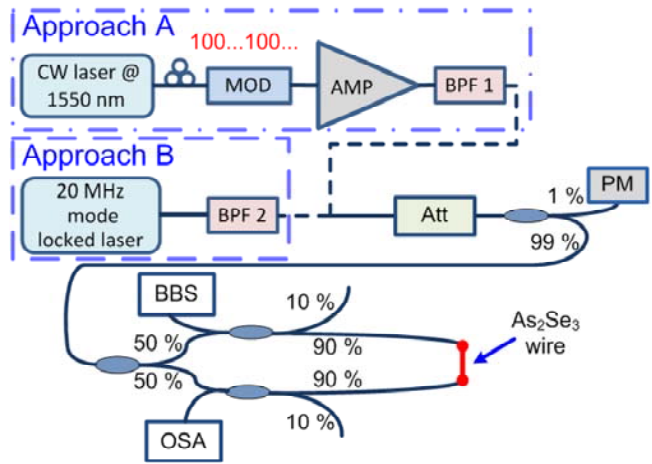

Figure 1. Experimental setup of approaches A and B to photo-induce a Bragg grating, Mod: Modulator, AMP: Amplifier, BPF: Bandpass filter, Att: Attenuator, BBS: Broadband optical source, OSA: Optical spectrum analyzer.

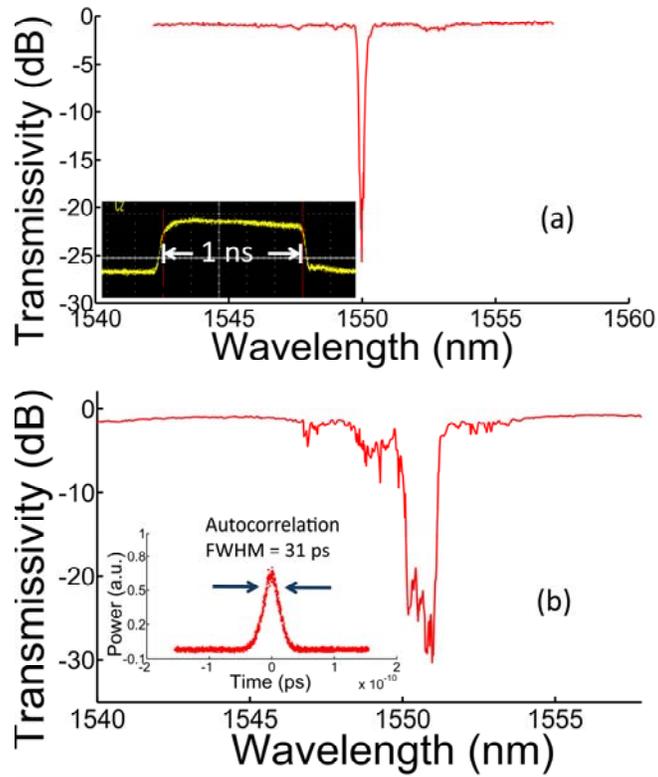

**Figure 2.** (a) Normalized transmission spectrum of a Bragg grating in microwire, written with 1 ns square pulses. (Inset) The pulse shape as observed on an oscilloscope (b) Normalized transmission spectrum of a Bragg grating written with ~22 ps Gaussian pulses. (Inset) Autocorrelation trace of the pulse.

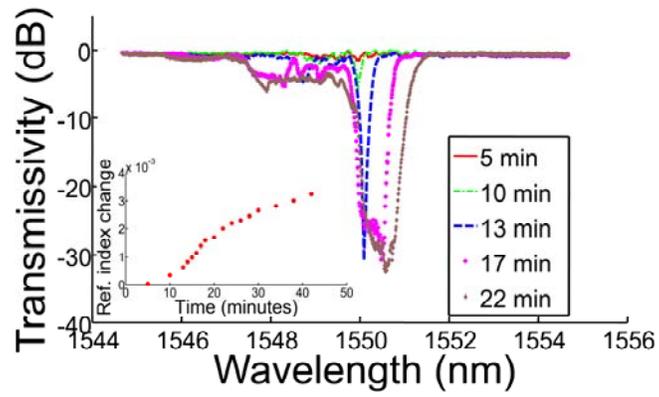

Figure 3. Transmission spectra of a Bragg grating as a function of time during photo-exposure. (Inset) Evolution of the AC refractive index change during the photo-exposure is also shown.

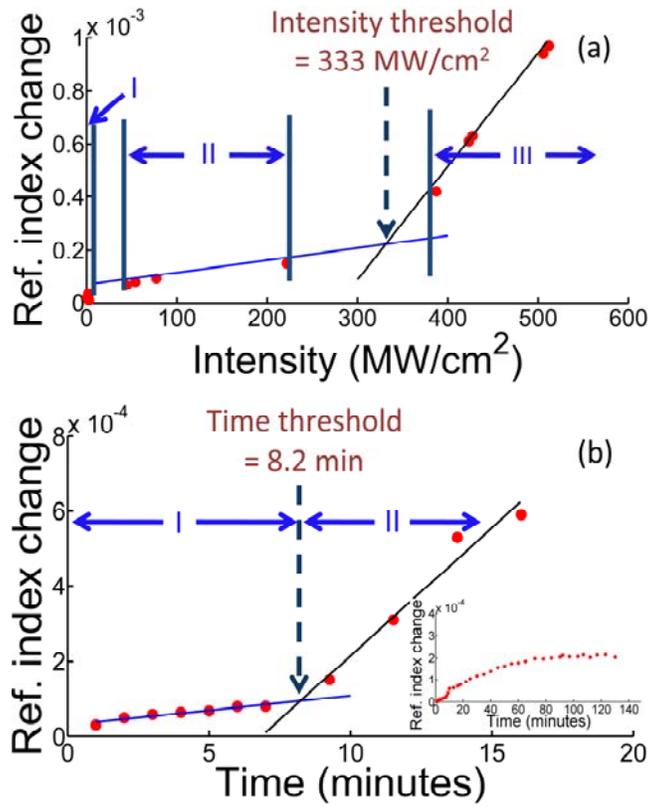

Figure 4. (a) Induced refractive index change as a function of input intensity for an exposure time of 20 minutes. Different intensity regions are labeled from I to III. (b) Induced refractive index change as a function of time for an input intensity sum of 425 MW/cm$^2$. The time scale is labeled in regions I and II on either side of the time threshold. (Inset) Induced refractive index change as a function of time for an input intensity sum of 52 MW/cm$^2$.

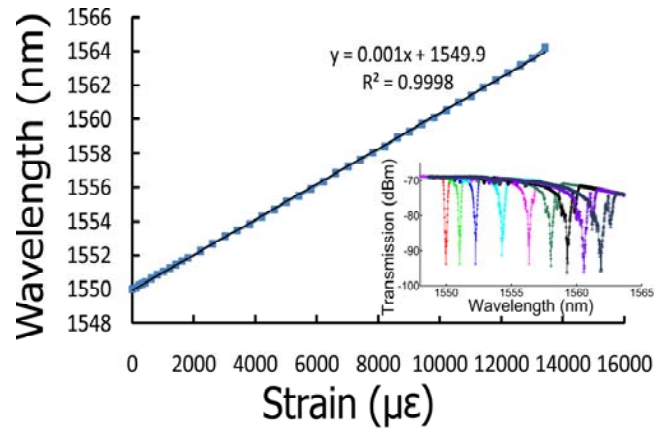

Figure 5. Strain measurement of a Bragg grating in an $As_2Se_3$ microwire. (Inset) Bragg grating transmission spectra for various strain values is shown as well.

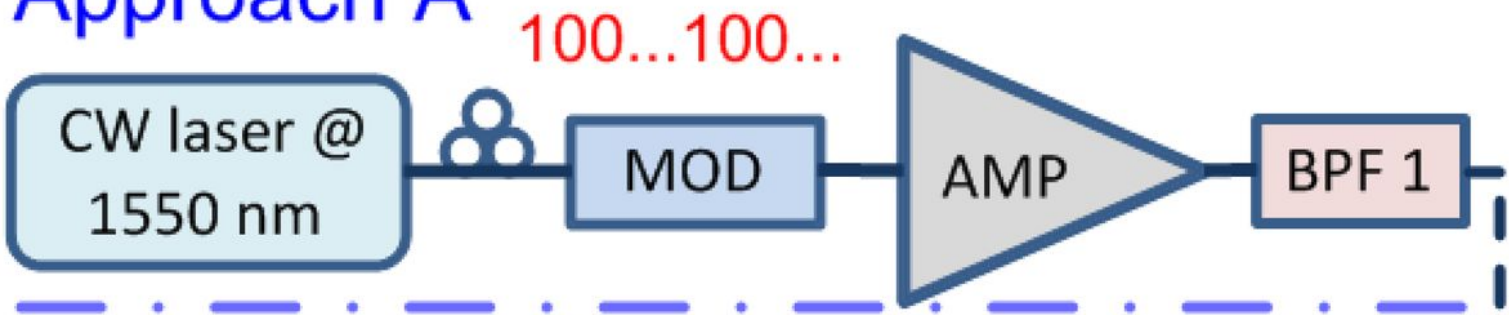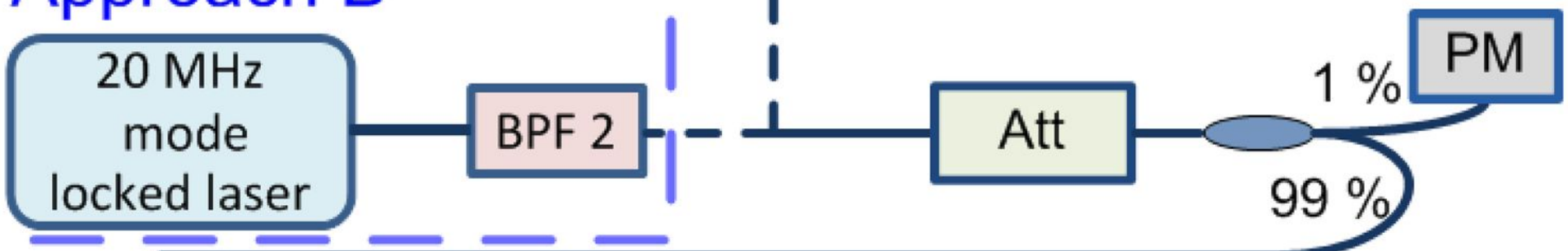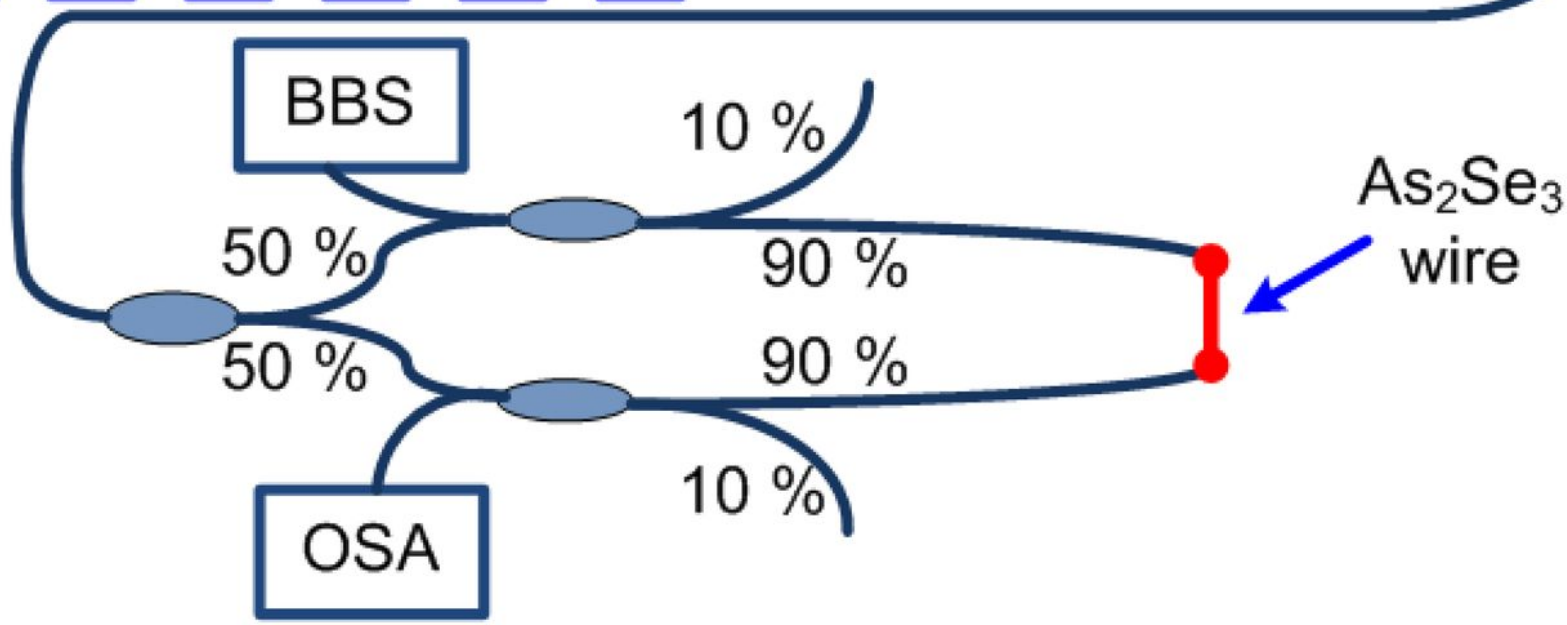

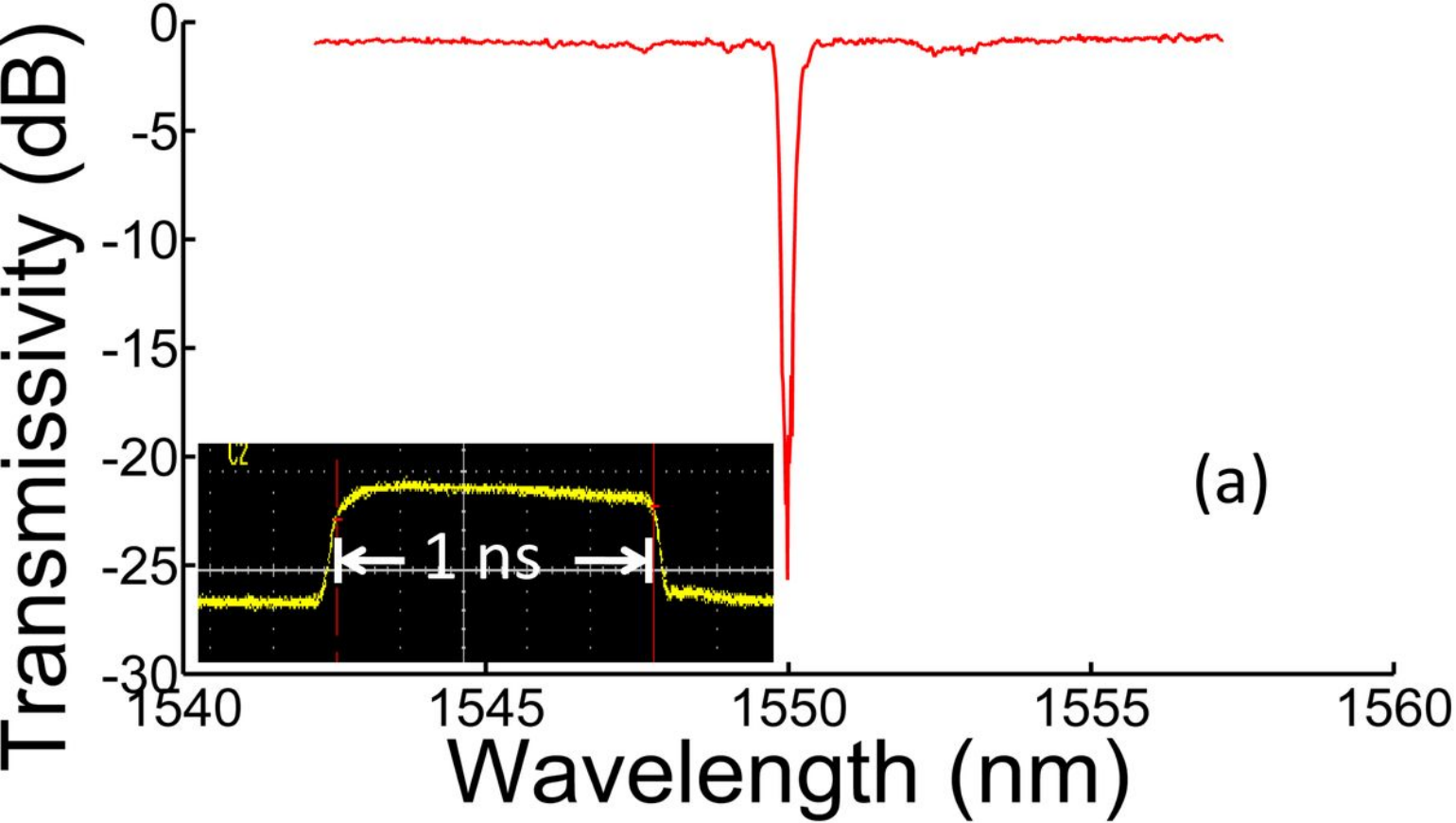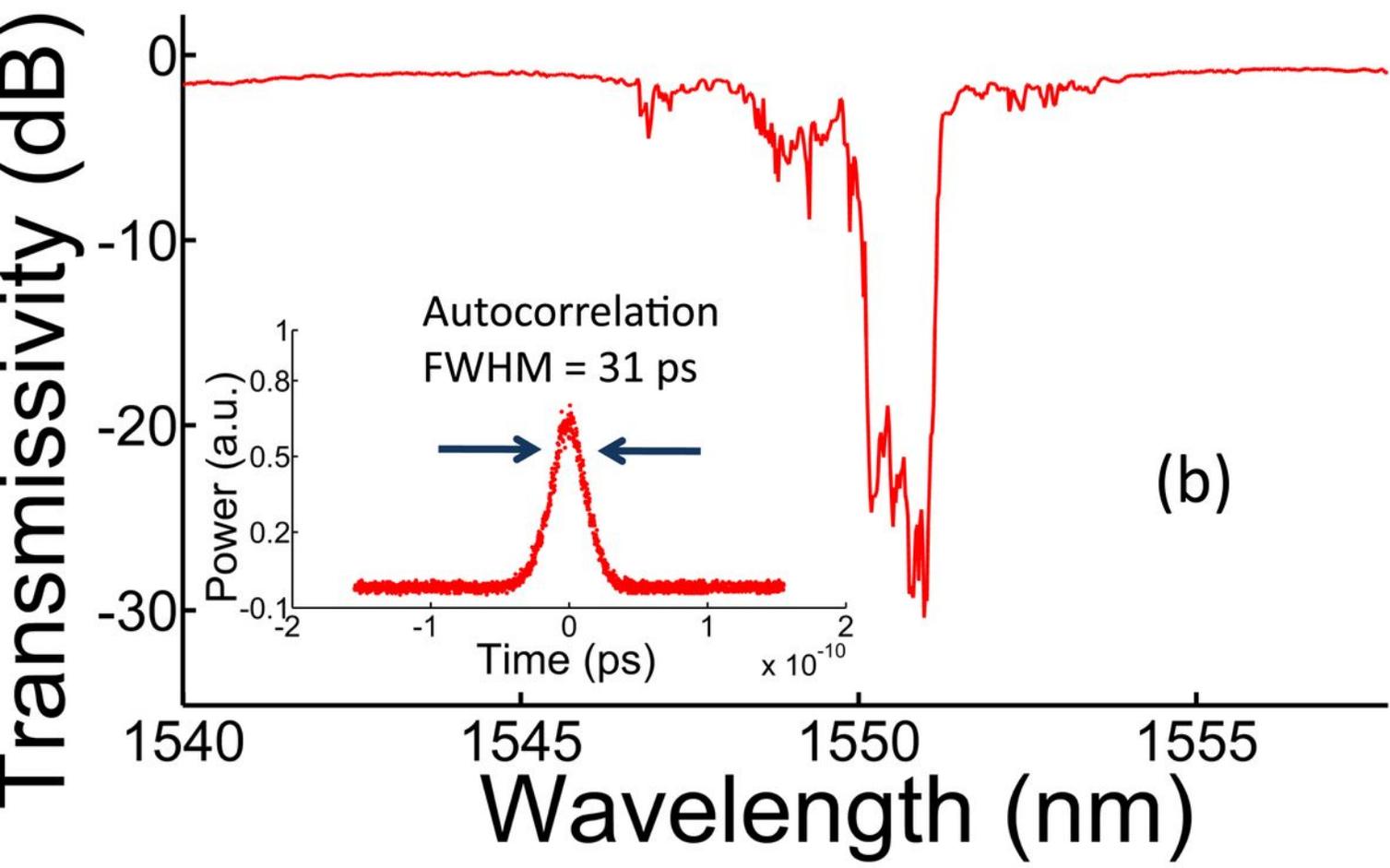

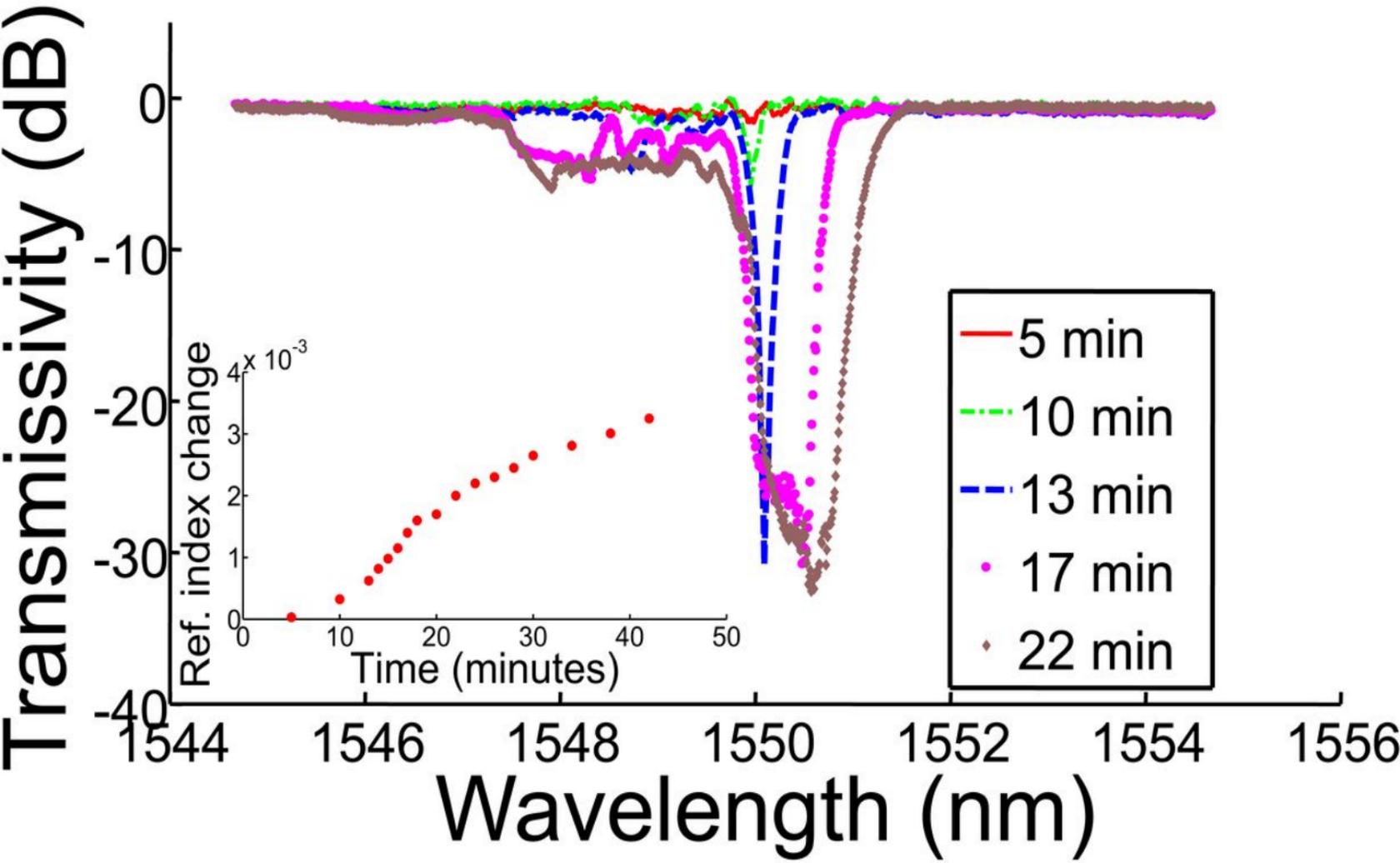

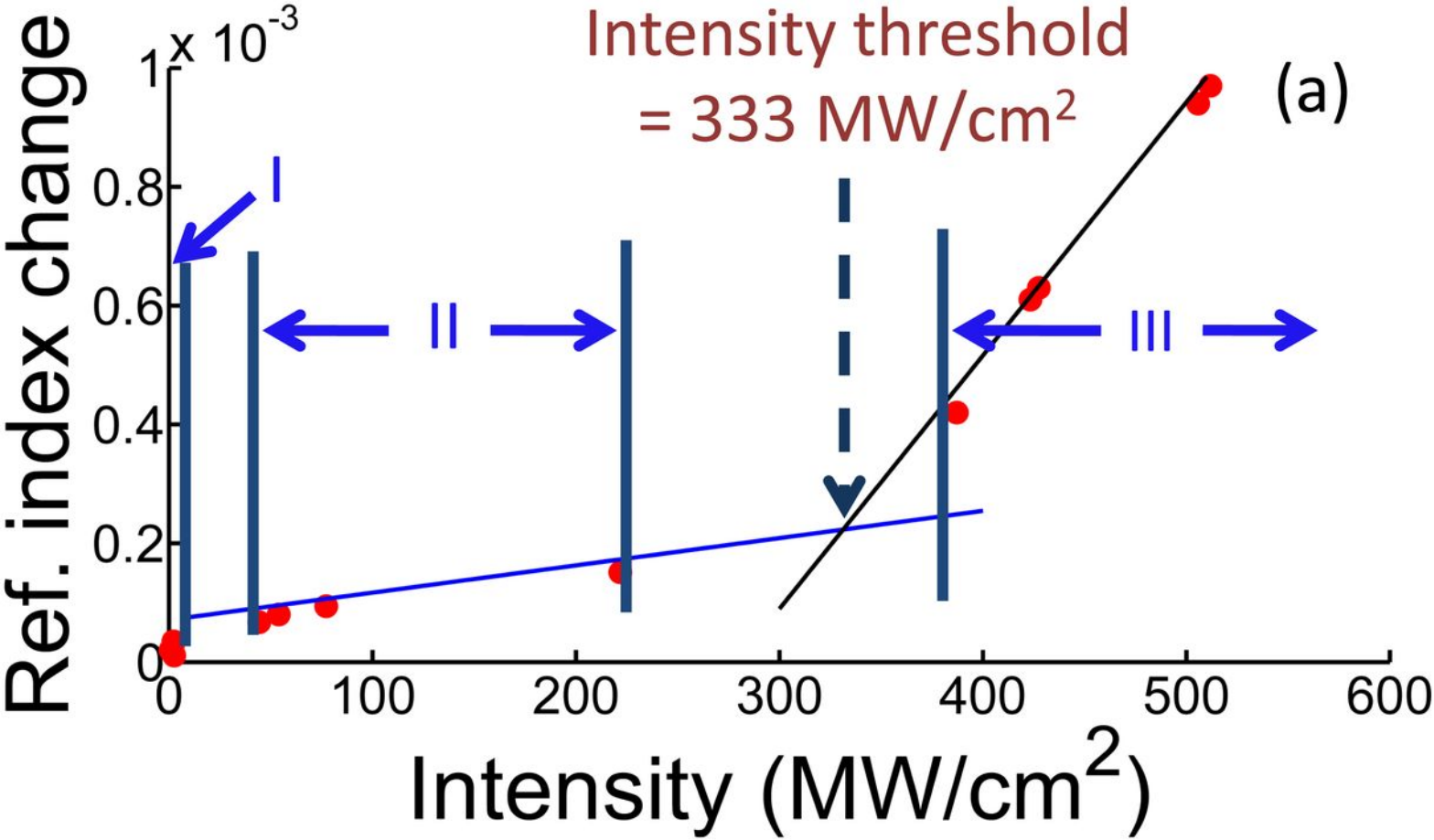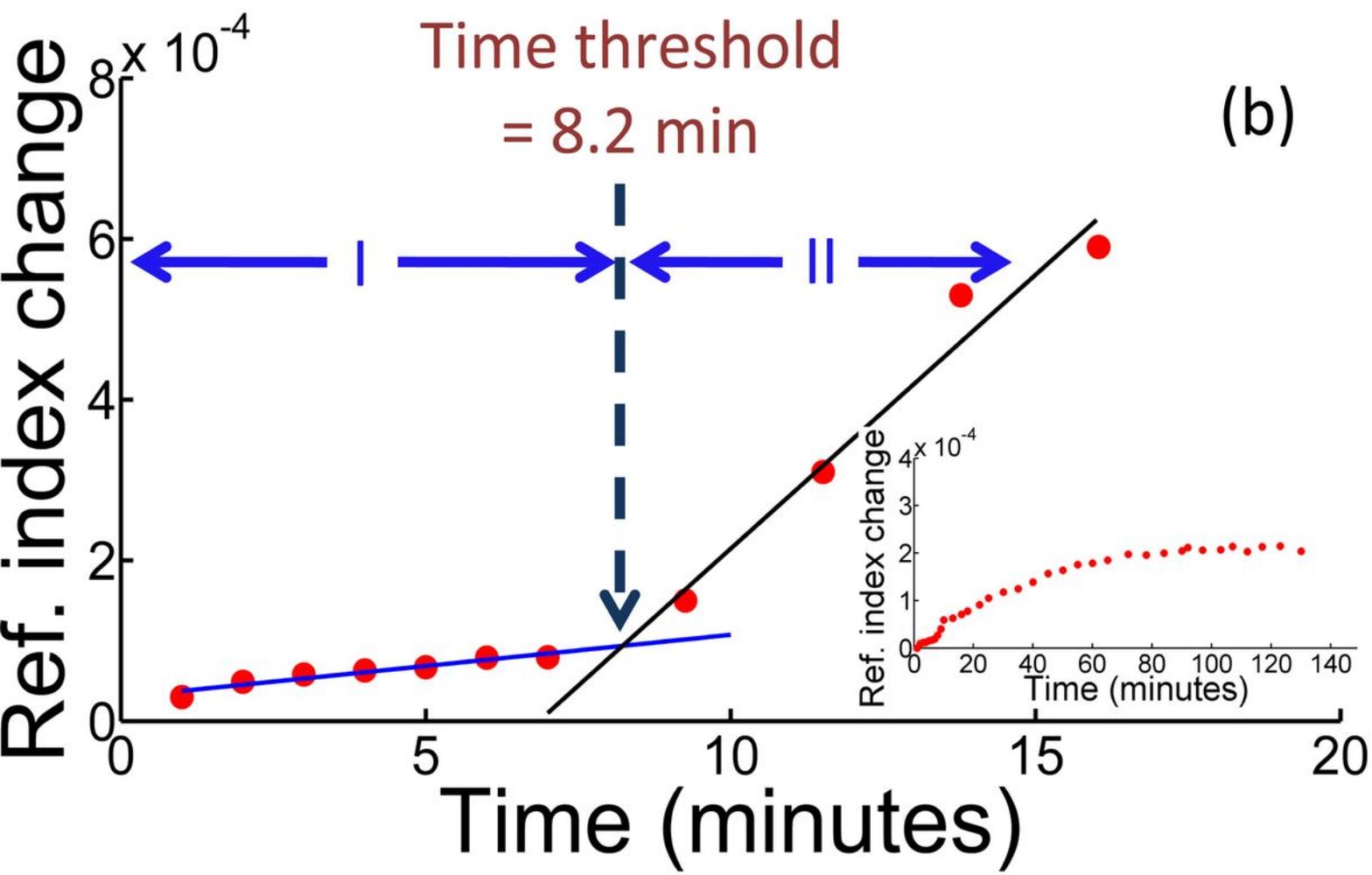

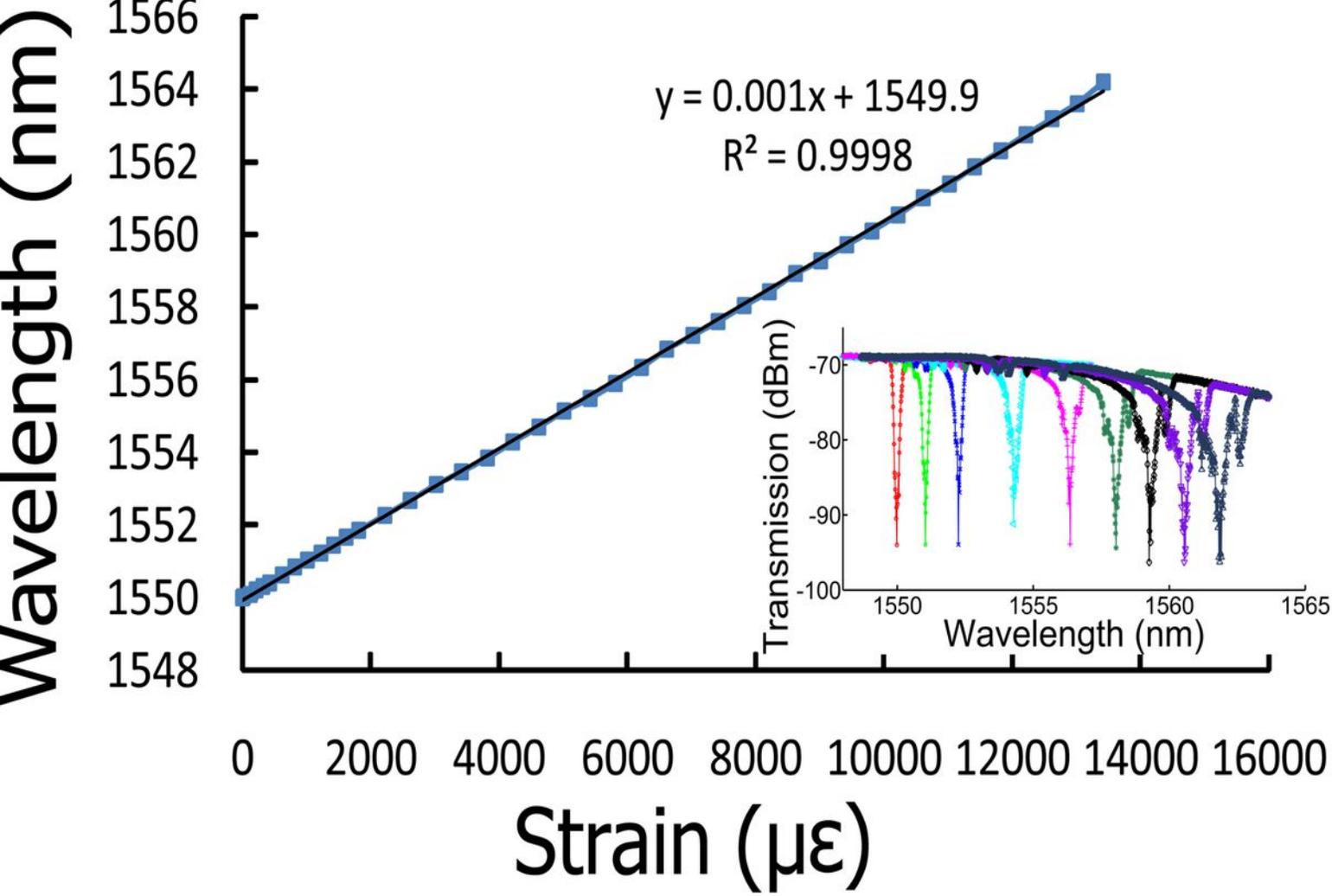